\documentclass{article}

\usepackage{PRIMEarxiv}

\usepackage[utf8]{inputenc} 
\usepackage[T1]{fontenc}    
\usepackage{hyperref}       
\usepackage{url}            
\usepackage{booktabs}       
\usepackage{amsfonts}       
\usepackage{nicefrac}       
\usepackage{microtype}      
\usepackage{lipsum}
\usepackage{graphicx}
\usepackage{cite}
\usepackage{amsmath,amssymb,amsfonts}
\usepackage{algorithmic}
\usepackage{graphicx}
\usepackage{textcomp}
\usepackage{framed,multirow}
\usepackage{booktabs}
\usepackage{multirow}
\usepackage{amsmath}
\usepackage{array}
\usepackage {color}
\newcolumntype{C}[1]{>{\centering\arraybackslash}p{#1}}

\usepackage{amssymb}
\usepackage{latexsym}
\usepackage{float}
\usepackage{enumitem}
\usepackage{caption}
\usepackage{tabularx}
\usepackage{booktabs}
\usepackage{pifont}
\usepackage{tabularx}
\usepackage{stfloats}
\usepackage{url}
\usepackage{hyperref}
\graphicspath{{media/}}     

\title{HA-HI: \\Synergising fMRI and DTI through Hierarchical Alignments and Hierarchical Interactions for Mild Cognitive Impairment Diagnosis}

\author{
  Xiongri Shen, Zhenxi Song\\
  Harbin Institute of Technology, Shenzhen\\
  Shenzhen\\
  \texttt{xiongrishen@stu.hit.edu.cn, songzhenxi@hit.edu.cn} \\
   \And
   Linling Li \\
  Shenzhen University\\
  Shenzhen\\
  \texttt{lilinling@szu.edu.cnl} \\
     \And
   Min Zhang\\
  Harbin Institute of Technology, Shenzhen\\
  Shenzhen\\
  \texttt{zhangmin2021@hit.edu.cn} \\
         \And
 Yichen Wei, Lingyan Liang\\
  The People’s Hospital of Guangxi Zhuang Autonomous Region\\
  Nanning\\
  \texttt{316644690@qq.com,lianglingyan163@126.com} \\
          \And
 Honghai Liu\\
  Harbin Institute of Technology, Shenzhen\\
  Shenzhen\\
  \texttt{honghai.liu@hit.edu.cn} \\
          \And
 Demao Deng\\
  The People’s Hospital of Guangxi Zhuang Autonomous Region\\
  Nanning\\
  \texttt{demaodeng@163.com} \\
            \And
 Zhiguo Zhang\\
 Honghai Liu\\
  Harbin Institute of Technology, Shenzhen\\
  \texttt{zhiguozhang@hit.edu.cn} \\
}

\begin{document}
\maketitle

\begin{abstract}
Early diagnosis of mild cognitive impairment (MCI) and subjective cognitive decline (SCD) utilizing multi-modal magnetic resonance imaging (MRI) is a pivotal area of research. While various regional and connectivity features from functional MRI (fMRI) and diffusion tensor imaging (DTI) have been employed to develop diagnosis models, most studies integrate these features without adequately addressing their alignment and interactions. This limits the potential to fully exploit the synergistic contributions of combined features and modalities. To solve this gap, our study introduces a novel \textbf{Hi}erarchical \textbf{A}lignments and \textbf{Hi}erarchical \textbf{I}nteractions (\textbf{HA-HI}) method for MCI and SCD classification, leveraging the combined strengths of fMRI and DTI. HA-HI efficiently learns significant MCI- or SCD-related regional and connectivity features by aligning various feature types and hierarchically maximizing their interactions. Furthermore, to enhance the interpretability of our approach, we have developed the Synergistic Activation Map (SAM) technique, revealing the critical brain regions and connections that are indicative of MCI/SCD. Comprehensive evaluations on the ADNI dataset and our self-collected data demonstrate that HA-HI outperforms other existing methods in diagnosing MCI and SCD, making it a potentially vital and interpretable tool for early detection. The implementation of this method is publicly accessible at \textcolor{blue}{https://github.com/ICI-BCI/Dual-MRI-HA-HI.git
}
\end{abstract}


\section{Introduction}
\label{sec:1}

Identifying Alzheimer's disease (AD) at an early stage, including mild cognitive impairment (MCI) and subjective cognitive decline (SCD), is crucial for the timely intervention \cite{cummings2020alzheimer, Nicholas2020MemoryCA}. In recent years, various image processing and pattern recognition methods have been developed to diagnose AD through brain imaging techniques, such as functional magnetic resonance imaging (fMRI) and diffusion tensor imaging (DTI) \cite{Oh2021LearnExplainReinforceCR, Janelidze2020PlasmaPI}. There is also much research to utilize various types of imaging features of fMRI and DTI to build machine learning and deep learning models for the diagnosis of MCI and SCD \cite{qiu2022multimodal,zhu2023multimodal,li2023alterations,liang2021recurrent}. However, most of the existing studies used relatively simple methods to analyze different features from different MRI modalities. For example, each MRI modality was independently processed to extract features, and different types of features were directly concatenated. Progression of AD causes complex, latent, and covariant structural and functional abnormalities in the human brain. It is vital to align features across different domains (regional or connectivity) as well as across different imaging modalities for a better understanding of the neuropathology of MCI and SCD and improved diagnostic accuracy \cite{zhang2023static,wang2022static}.


Current methods for analyzing multi-modal MRI data are technically categorized into three primary categories: feature engineering, deep learning, and their integrative approaches \cite{jeon2020enriched,dong2021integration}. Feature engineering depends on domain knowledge and often encounters sparsity due to feature choices, whereas deep learning enables an end-to-end pipeline but necessitates dedicated forward propagation designs for MRI synergy exploitation.

\begin{figure*}[t]
    \centering
    \includegraphics[scale=.29]{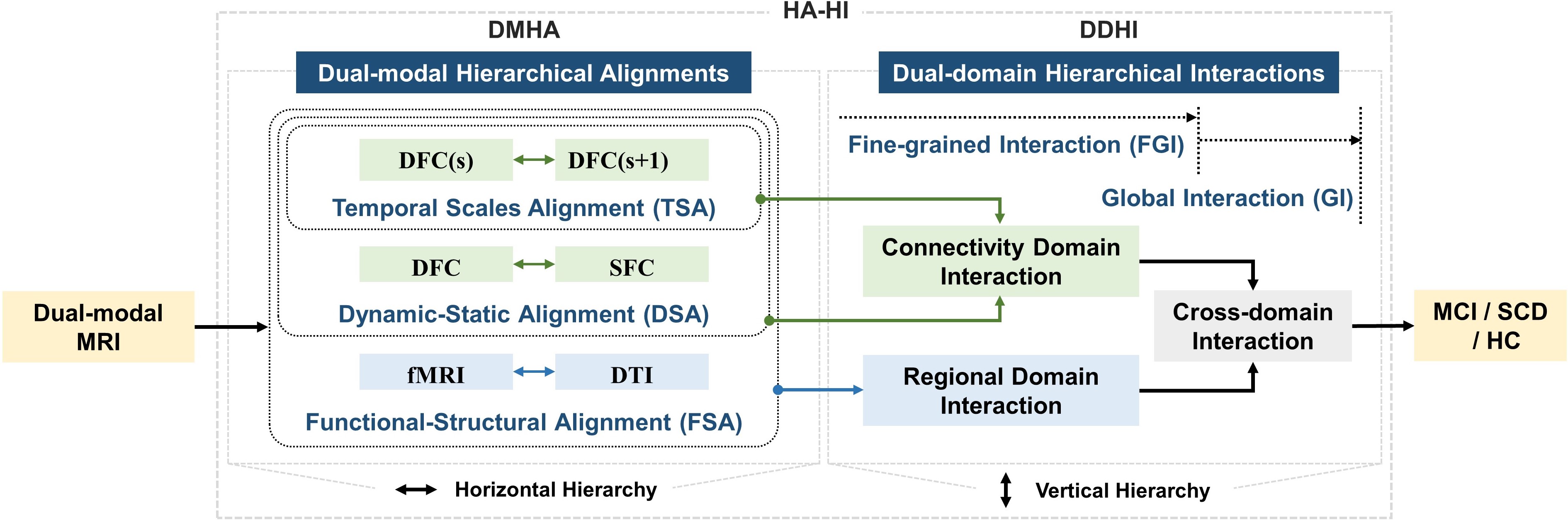}
    \caption{Workflow of \textbf{\textit{HA-HI}}: This framework enhances cognitive impairment identification via dual-modal hierarchical alignments (\textit{DMHA}) and dual-domain hierarchical interactions (\textit{DDHI}). \textit{DMHA} aligns diverse features from fMRI and DTI modalities horizontally, and \textit{DDHI} optimizes feature fusion across  regional and connectivity domains vertically.}
    \label{fig: Fig1}
\end{figure*}

In terms of MRI features, these methods primarily focus on regional and connectivity domains. Neural degenerations due to cognitive decline accumulate within brain regions and are detectable through neuromarkers measured by fMRI and DTI techniques \cite{yang2018gradual,schmithorst2005cognitive}. Notably, changes in fractional anisotropy (FA) in the brain's white matter, particularly in the frontal and occipital lobes, are indicative of neurodegenerative diseases \cite{schmithorst2005cognitive}. Accordingly, FA values from DTI were often used to illustrate structural anomalies. On the other hand, functional regional activities were often characterized by the brain's spontaneous activity intensity, for example, measured by the Amplitude of Low-Frequency Fluctuations (ALFF) in fMRI. Prior research indicates that variations in ALFF values are associated with cognitive decline, specifically in the hippocampus and thalamus \cite{yang2018gradual}. Connectivity features are also of great importance in the progression of AD. fMRI-based functional connectivity, including both static (SFC) and dynamic (DFC) aspects, has been extensively investigated for differentiating SCD and MCI from normal status \cite{Srikanthanathan2021DifferencesIF,xue2021disrupted,dautricourt2022dynamic}. 
These studies indicate distinct connectivity patterns in specific networks, such as the default mode network (DMN) and frontal-parietal network (FPN), under varying cognitive conditions. 


However, existing methods have not fully synergized the rich, complementary information across various MRI modalities and features. Specifically, many studies overlook co-variations or relationships among MRI features in their diagnostic model designs.

\begin{enumerate}
    \item \textbf{Functional-structural co-variation:} Regional neuromarkers in functional (from fMRI) and structural (from DTI) perspectives are inter-dependent and covaried \cite{xu2021different,avila2022brain,zhong2022shared}, which calls for new methods to achieve functional-structural alignments. 
    \item \textbf{Relationship between SFC and DFC:} Both SFC and DFC features are found to be indicative of MCI/SCD. But most existing SFC or DFC research overlooks the comprehensive exploration of a learning-based integration of SFC and DFC aspects, which necessitates considering their inter-correlated nature \cite{wang2022white, Kam2020DeepLO, Briend2020AberrantSA}. Thus, combining dynamic with static FC features could unlock intrinsic network properties.
    \item \textbf{Multi-scale DFC integration:} Current DFC methods often rely on fixed window sizes \cite{duda2021validating,zhang2022test}, potentially affecting retest reliability. Prior work with high-temporal resolution EEG data suggests cognitive impairment significantly alters connectivity at multiple scales \cite{8356590}. Thus, multi-scale temporal integration of DFC is desired.
    \item \textbf{Regional-connectivity coupling:}  Regional anomalies are also intertwined with connectivity abnormalities in several neurological disorders, so regional-connectivity coupling/de-coupling can provide meaningful neuromarkers for MCI and SCD \cite{lei2021auto,wei2022functional}. Hence, regional features and connectivity features should also be integrated to better predict MCI/SCD.
\end{enumerate}

In a nutshell, existing studies lack a systematic amalgamation of these diversified features across both functional and structural levels in the regional and connectivity domains.


To address existing challenges, this paper introduces a novel hierarchical framework, named \textbf{H}ierarchical \textbf{A}lignments and \textbf{H}ierarchical \textbf{I}nteractions (\textbf{\textit{HA-HI}}), for identifying MCI, SCD, and healthy people using fMRI and DTI.
As depicted in Fig.\ref{fig: Fig1}, \textbf{\textit{HA-HI}} advances fMRI and DTI co-analysis by 
(1) implementing Dual-Modal (fMRI and DTI ) Hierarchical Alignments (\textit{DMHA}) to synchronize DFC across various time scales, bridge static and dynamic connectivity patterns, and align regional functional and structural abnormalities; 
(2) incorporating Dual-Domain (regional and connectivity) Hierarchical Interactions (\textit{DDHI}) to integrate features across both regional and connectivity domains, ranging from fine-grained to global levels.
Furthermore, given that \textbf{\textit{HA-HI}} capitalizes on dual-modal MRI synergies, it is crucial to develop an interpretive technique for this 'black-box' model. Thereby, the primary contributions of this study include:
\begin{enumerate}[label=$\bullet$]
    \setlength\itemsep{0em}
    \setlength\parsep{0em}
    \item Development of \textbf{\textit{HA-HI}}, a learning-based hierarchical framework for comprehensive dual-modal MRI analysis.
    \item Validation of the \textbf{\textit{HA-HI}} method's effectiveness in diagnosing MCI and SCD is demonstrated through two datasets and an accompanying ablation study.
    \item Introducing the innovative Synergistic Activation Mapping (\textit{SAM}) technique for a quantitative and qualitative evaluation of dual-modal MRI synergy effects, facilitating the inference of significant connectivities and regions within the trained \textbf{\textit{HA-HI}}.
\end{enumerate}

\textit{Note: Standard abbreviations and specialized terms are in }UPPERCASE \textit{regular font, while new method abbreviations are in UPPERCASE italics, with \textbf{HA-HI} emphasized in bold.}

\begin{figure*}[t]
    \centering
    \includegraphics[scale=.29]{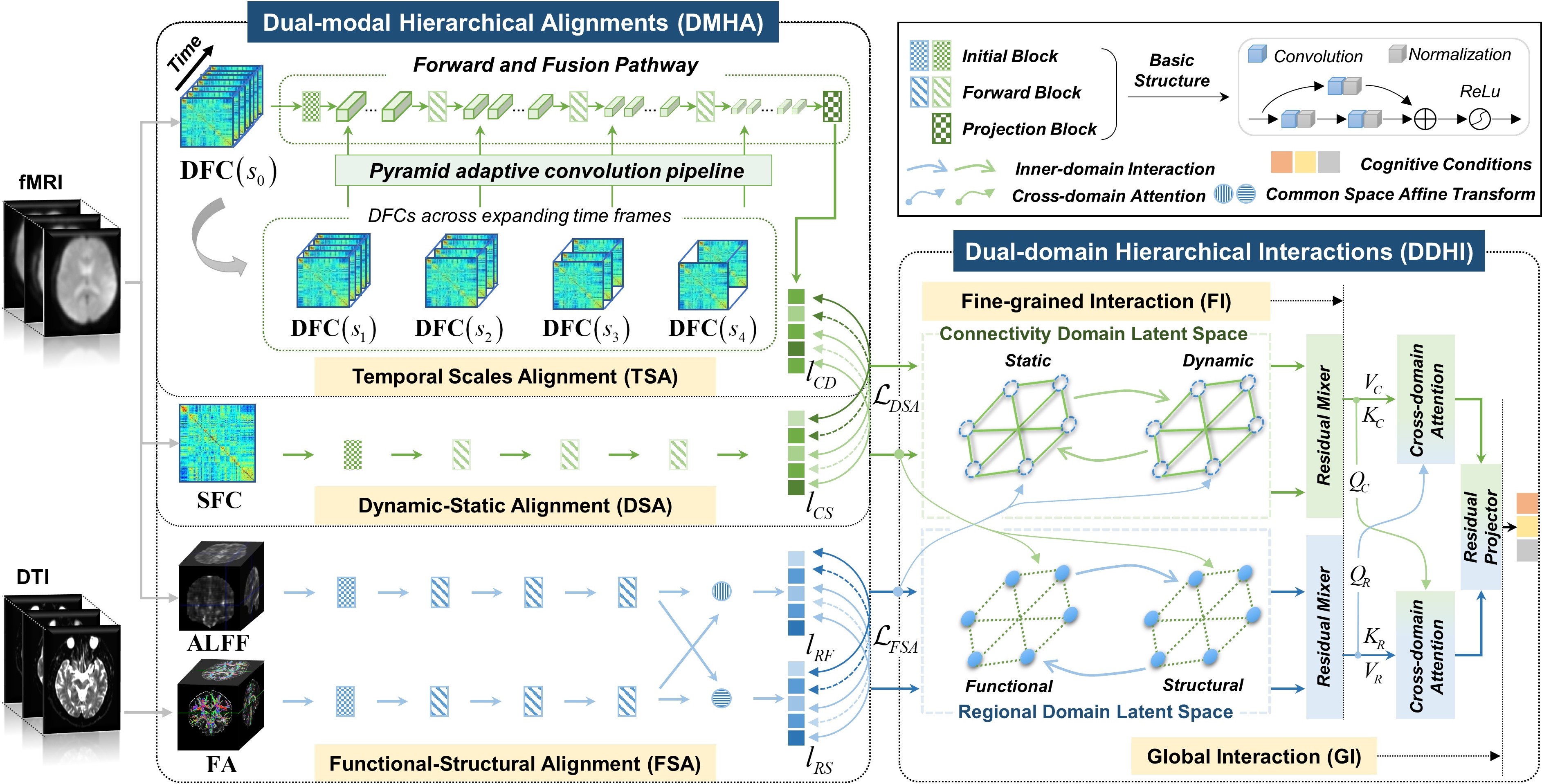}
    \caption{The technical details of \textbf{\textit{HA-HI}}, developed for cognitive impairment detection using fMRI and DTI inputs, capitalize on \textit{DMHA}'s strengths in performing hierarchical alignments between dynamic temporal scales, integrating dynamic and static networks, and correlating functional with structural features, and on \textit{DDHI}'s role in conducting hierarchical interactions from the fine-grained to the global level, to fuse features across regional and connectivity domains.
    }
    \label{fig: Fig2}
\end{figure*}

\section{Method}
\label{sec:2}

The framework of \textbf{\textit{HA-HI}} comprises \textit{DMHA} and \textit{DDHI} modules (Fig. \ref{fig: Fig2}). 
In the \textbf{\textit{HA-HI}} framework, \textit{DMHA} employs a  horizontal hierarchical structure to (1) harmonize various temporal scales in DFC, (2) align DFC and SFC patterns, and (3) integrate regional functional and structural features  (\textit{refer to Section \ref{sec:2.1}}). 
Concurrently, \textit{DDHI} implements a vertical hierarchical structure, facilitating interactions through attention mechanisms that evolve from fine-grained to global levels (\textit{refer to Section \ref{sec:2.2}}).
Moreover, we introduced a \textit{SAM} technique to interpret the significantly affected brain networks and regions. (\textit{refer to Section \ref{sec:2.3}}).
Specifically, within the encoding pathway of dual-modal MRI, the basic structure utilizes convolutions, featuring skip connections. These connections mitigate the issues of gradient vanishing and explosion and enable the model to effectively learn representative mappings, thereby enhancing overall performance. 

\vspace{-3mm}
\subsection{Dual-Modal Hierarchical Alignments - DMHA
\label{sec:2.1}}

\subsubsection{Temporal Scale Alignment (TSA)
\label{sec:2.1.1}
}


Temporal scale alignment aims to conduct a co-analysis of multi-scale DFCs with a reasonable alignment of different temporal scales. First, the multi-scale DFCs are generated using the spatial pyramid pooling mechanism \cite{Fang2021multi-organ,xing2022multi-scale}, and the functional connectivity is estimated by Pearson's correlation coefficients between different brain regions. 
The dimensions of DFC are expressed as \( W \times H \times D_s \), where \( D_s = \max(k) \) is determined by both the scale factor \( s \) and the total number of dynamic sampling points, denoted as \( T \), measured at the original temporal scale in DFC. In the forward propagation, the dimensions of the feature maps, extracted from multi-scale DFCs, are downsampled to \(W \cdot 2^{-l} \times H \cdot 2^{-l} \times D_s \cdot 4^{-l}\), where \( l \) indicates the level of the convolutional block in the encoding pathway. Consequently, the selected scale factor \( s \) is configured according to the encoding level as \(s = 2^l \cdot \delta \), where \( \delta \) denotes the magnification factor of the multi-scale DFCs in the temporal dimension.

\begin{figure}[b]
    \centering 
    \includegraphics[scale=.35]{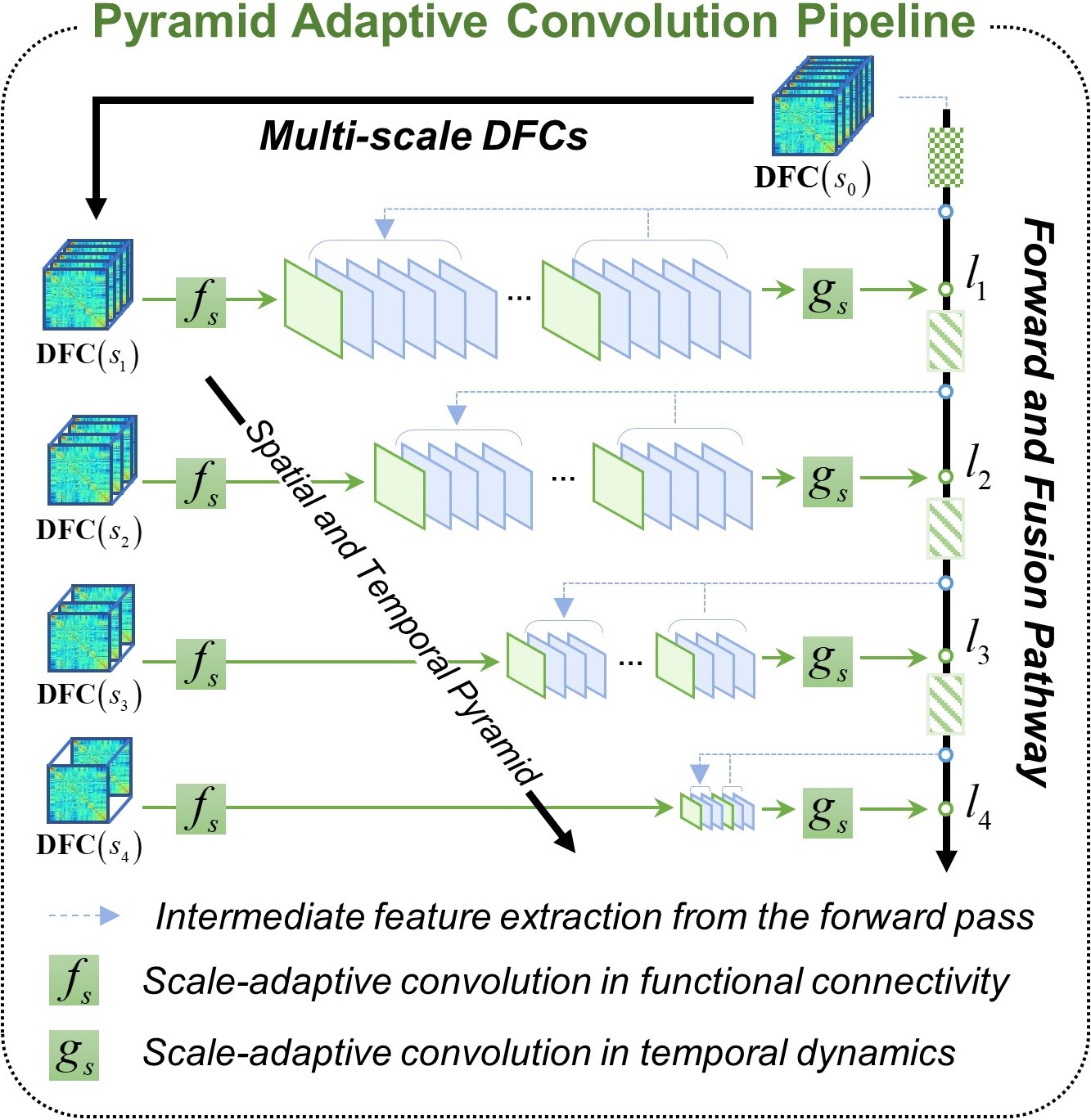}
    \caption{Strategy for Multiscale DFC alignment.}
    \label{fig: Fig3}
\end{figure}

To address the pyramid structures present in both spatial dimensions and temporal scales, we introduce a pyramid adaptive convolution pipeline (Fig. \ref{fig: Fig3}), which aligns the multi-scale DFCs derived at various temporal scales. To ensure logical consistency during the extraction of feature flows from DFCs of different scales, we have designed two decoding modules, \(f_s(\cdot)\) and \(g_s(\cdot)\). These modules follow the architecture of the forward block depicted in Fig. \ref{fig: Fig1} and exhibit scalability across various dimensions. The first module, \(f_s(\cdot)\), adaptively adjusts the spatial dimensions of features and explores the functional connectivity patterns. The second module, \(g_s(\cdot)\), adaptively aligns the temporal resolution of features, synthesizing the temporal dynamics from multi-scale features.

More specifically, \(DFC(s)\) first passes through the convolutional module \(f_s(\cdot)\), transforming the dimension \( W \times H\) to conform with the intermediate feature maps extracted from the forward encoding pathway at level \(l = \log_2\left(\frac{s}{\delta}\right)\). Subsequently, these maps are concatenated along the depth dimension utilizing the equal interval insertion method, thereby aligning the temporal resolution. The merged feature maps, having a depth of \(D=D_s \cdot 4^{-l}+T \cdot 4^{-l}\), undergo further transformations via the depth-wise convolutional module \(g_s(\cdot)\), ensuring synchronization of the depth with the forward pathway.

\subsubsection{Dynamic-Static Alignment (DSA)
\label{sec:2.1.2}
}


Representations learned from DFC unveil the connectivity attributes of the dynamic brains \cite{hutchison2013dynamic}. Conversely, features from SFC capture the essence of static brain networks. We employ an encoding pathway, consisting of four convolutional blocks, which mirrors the structures used for DFC analysis, to convert SFC data into high-level embeddings. For consistent inference, the primary embeddings—encoding the cognitive level—extracted from SFC and DFC must be both indicative and analogous. Motivated by the principles of contrastive learning \cite{zhu2023cl}, 
we introduced a contrastive loss to ensure alignment between dynamic and static representations. Denoting the high-level dynamic and static embeddings as \(Z_d\) and \(Z_s\) respectively, the dynamic-static contrastive loss \(\mathcal{L}_{DSA}\) for the positive pairs originating from the same sample is
    \begin{equation}
        \mathcal{L}_{DSA} = - \log{\frac{exp\left(sim(Z_d,Z_s)/\tau\right)}{\sum^{2N}_{\mathrm{j}=1}1_{[\mathrm{j} \neq \mathrm{d}]}exp\left(sim(Z_d,Z_\mathrm{j})/\tau\right)}},
        \label{eq: L_DSA}
    \end{equation}
    \begin{equation}
        {sim}(Z_d, Z_s) = \frac{Z_d \cdot Z_s}{\|Z_d\|_2 \cdot \|Z_s\|_2},
        \label{eq: SIM}
    \end{equation}
where \(\tau\) is a temperature parameter, \(exp\) signifies the exponential function, \(1_{[j \neq d]}\) denotes the indicator function, and \(N\) corresponds to the number of samples in a mini-batch. Here, \(sim\) implies a similarity metric, such as the cosine similarity. For this study, we adopted the similarity measure presented in Eq.(\ref{eq: SIM}) to synchronize the paired embeddings derived from both dynamic and static scenarios.

\subsubsection{Functional-Structural Alignment (FSA)
\label{sec:2.1.3}
}


As stated in the Introduction, the observed commonalities and  disparities in deteriorating regions between functional and structural brain states led us to develop a two-step strategy for functional-structural alignment:

\begin{enumerate}[label=(\arabic*)]
    \item
    Inspired by latent space learning \cite{chen2023orthogonal}, the first step integrates functional features (\(z_F\)) and structural features (\(z_S\)) within a shared space. This uncovers cognition-centric information invariant across functional-structural common spaces defined by various affine transformations. In our tests, two simple transformations, as shown in Eq.(\ref{eq: CommonSpace}), proved sufficiently effective.
    \begin{equation}
        Z_{*} = z_F \times z_S, \qquad Z_{+} = z_F + z_S
        \label{eq: CommonSpace}
    \end{equation}
    \item
    The subsequent step exploits contrastive learning, drawing parallels with the dynamic-static alignment as detailed in Eqs. (\ref{eq: L_DSA}) and (\ref{eq: SIM}). In this phase, embeddings within the functional-structural common space—originating from the affine transformations \(Z_*\) and \(Z_+\)—are aligned using the contrastive loss \(\mathcal{L}_{FSA}\) (Eq. (\ref{eq: L_FSA}). The ultimate goal is to identify distinct features that maintain consistency across both combination strategies.
     \begin{equation}
        \mathcal{L}_{FSA} = - \log{\frac{exp\left(sim(Z_*,Z_+)/\tau\right)}{\sum^{2N}_{\mathrm{i}=1}1_{[\mathrm{i} \neq \mathrm{*}]}exp\left(sim(Z_*,Z_\mathrm{i})/\tau\right)}}
        \label{eq: L_FSA}
    \end{equation}
\end{enumerate}

\subsection{Dual-Domain Hierarchical Interactions - DDHI}
\label{sec:2.2}


To overcome challenges within dual-modal networks, such as mitigating overfitting caused by a large parameter space and effectively managing abundant information \cite{wang2020makes}, we designed a dual-domain hierarchical interaction. This approach leverages complementary patterns across modalities, guided by robust mathematical logic and a neuroscience perspective. Specifically, this mechanism employs select data aspects to hierarchically direct attention, rather than indiscriminately aggregating information from all available data sources. We've defined latent representations in the dynamic and static connectivity domains as \(l_{CD}\) and \(l_{CS}\), respectively, while those in the functional and structural regional domains are denoted as \(l_{RF}\) and \(l_{RS}\). The mechanism can be divided into two hierarchical levels: fine-grained interaction with modulated attention (see Section \ref{sec:2.2.1}), and global interaction facilitated by cross-domain attention (see Section \ref{sec:2.2.2}).

\subsubsection{Fine-grained Interaction with Modulated Attention (FI)}
\label{sec:2.2.1}


In the latent space of the connectivity domain, only \(l_{RF}\) from the regional domain is employed for interactions with \(l_{CD}\) and \(l_{CS}\). This choice highlights their modality homogeneity, which is rooted in fMRI. In the regional domain's latent space, only \(l_{CS}\) from the connectivity domain is utilized for \(l_{RF}\) and \(l_{RS}\). This selection depends on their uniform representation of the brain's steady state from a static perspective, compared to dynamic interpretations. Given this, we defined four interactions applicable to \(l_{CD}\), \(l_{CS}\), \(l_{RF}\), and \(l_{RS}\) as
    \begin{equation}
        I_{\text{FG}}\left(l_{*}, l_{\mathcal{o}}, l_{+}\right) = \langle \sigma\left(\frac{\langle l_{*}, l_{\mathcal{o}} \rangle}{\sqrt{\lambda}} \right), l_{+}\rangle
        \label{eq: InterInner},
    \end{equation}
where \(l_{*}\) denotes either \(l_{RF}\) or \(l_{CS}\), which are used to compute the modulated attention. These weights calibrate the inner-domain interaction for the connectivity or regional domains, respectively. Meanwhile, \(l_{\mathcal{o}}\) and \(l_{+}\) represent two types of embeddings obtained within the same domain. Importantly, \(\sqrt{\lambda}\) scales the large magnitudes produced by the dot products, denoted by \(\langle \cdot \rangle\), ensuring that the Softmax function, represented by \(\sigma\), avoids the vanishing gradient issue.

\subsubsection{Global Interaction with Cross-domain Attention (GI)}
\label{sec:2.2.2}


By assigning the output of the embedding from the residual mixer as the Query, Key, and Value vectors, the attention mechanism is inspired by \cite{Wei_2020_CVPR} and formulated as
    \begin{equation}
        I_{\text{G}_{+}} = \sigma\left(\frac{Q_{*} \cdot K_{+}^T}{\sqrt{d_k}}\right)V_{+},
    \end{equation}
where the symbols \(*\) and \(+\) indicate connectivity or regional domains, respectively. Here, the Query vector \(Q_{*}\) from one domain interacts with the Key vector \(K_{+}\) from another domain through dot product, seeking the global relationships in dual-modal MRI from regional and connectivity views. Similarly, such a relationship is normalized by the Softmax function (symbolized as \(\sigma\)) to produce the soft attention matrix and rescaled by a scaling factor \(\sqrt{d_k}\) to regulate magnitude order. The Value vector \(V_{+}\) is finally weighted to generate integrated features \(\text{G}_{+}\) for each domain.

To accomplish the mapping to cognitive conditions (e.g., MCI, SCD, etc.), the residual projector fuses the embeddings \(\text{G}_{C}\) and \(\text{G}_{R}\) (C: Connectivity; R: Regional) and reduces the dimension to match the number of classes. Cross entropy is used for estimating the classification loss $\mathcal{L}_{CLS}$. Consequently, for the back-propagation process, we utilized the summarized loss as defined in Eq. (\ref{eq: Loss}).
    \begin{equation}
    \label{eq: Loss}
        \mathcal{L} = \mathcal{L}_{CLS} + \mathcal{L}_{DSA} +  \mathcal{L}_{FSA}
    \end{equation}


\vspace{-6mm}
\subsection{Synergistic Activation Mapping - SAM}
\label{sec:2.3}



Our proposed \textbf{\textit{HI-AI}} architecture, which capitalizes on hierarchical MRI synergies, has led us to develop a visualization technique named \textit{SAM}. This method is vital for clinical applications as it elucidates the synergistic contributions from different modalities and enhances the process of determining cognitive conditions using dual-modal MRI data. \textit{SAM} is grounded in the principles of Score-CAM \cite{wang2020score}, originally intended for convolutional neural networks (CNNs) in computer vision tasks. Differently, \textit{SAM} diverges by being adapted for hybrid models, specifically to handle three-dimensional multi-modal MRI features through parallel feed-forward propagation, independent of global average pooling layer reconstruction and gradient calculations.

In detail, the \textit{SAM} method estimates the importance of brain regions by considering both the absolute values of the original multi-modal MRI (denoted as \(R^i\)) and the relative significance of neurons highlighted in the feature map (denoted as \(F^i\)), where \(i\) denotes the \(i\)\textit{th} modality. To ensure the explanations focus on the contribution of each modality without confounding factors, we identified the learned high-level features, before feeding them into hierarchical interactions within the HI-AI framework, as \(F^i\). Consequently, \(F^i=f_{DMHA}(I^i)\), where \(f_{DMHA}\) signifies the mapping from input to the extracted layer, and \(I^i\) symbolizes dual-modal MRI features, such as DFC, SFC, ALFF, and FA. For the patient class \(c\), the explainable results produced by \textit{SAM} can be defined as follows:
    \begin{align}
        SAM_c(I) = \Bigg\{ &\text{ReLU}\left( \Upsilon\left( \sum_{k=1}^{K} I^{i}_{k} \cdot M^{i}_{k} \right) \right) \mid i = 1, \ldots, M \Bigg\}
    \end{align}
    \begin{equation}
        I^{i}_{k} = \text{Softmax}\left[ f_{\text{MMHA}}\left( I^{i} + \Upsilon(M^{i}_{k}), c\right) \right]
    \end{equation}
    \begin{equation}
        M^{i}_{k} = \mathcal{N}(F^{i}_{k})
    \end{equation}

\vspace{-3mm}
Here, \(\mathcal{N}(\cdot)\) is a pre-defined function used for normalization, aligned with the batch-normalization technique applied to the input. \(M^{i}_{k}\) indicates the relative importance of regions in the \(k\)\textit{th} feature map of the \(i\)\textit{th} modality. \(\Upsilon\) represents trilinear interpolation that calibrates the dimensions of \(M^{i}_{k}\) to match \(I^{i}\), enabling the integration of absolute values in dual-modal MRI features with the relative significance of neurons highlighted in the feature map. Subsequently, these integrated features are delivered to the mapping \(f_{DMHA}\) to generate weights for each cognitive condition \(c\). A softmax layer is applied to produce a probability map, which is used to weigh the feature map \(M^{i}_{k}\) via Hadamard multiplication. The weighted map is then upsampled through \(\Upsilon\) and finally activated by the \(\text{ReLU}(\cdot)\) function, ensuring a non-negative activation map.

\section{Datasets Description}
\label{sec:3}

To validate the generalizability of \textbf{\textit{HA-HI}}, we utilized resting-state fMRI and DTI data from two distinct sources: \textbf{(1)} Data collected from the hospital, the First Affiliated Hospital of Guangxi University of Traditional Chinese Medicine, hereafter referred to as the 'GUTCM Dataset', and \textbf{(2)} The Alzheimer's Disease Neuroimaging Initiative (ADNI) repository \cite{petersen2010alzheimer}. Descriptions of the GUTCM dataset (Section \ref{sec:3.1}) and the ADNI dataset (Section \ref{sec:3.2}) 

\vspace{-3mm}
\subsection{Dataset from the hospital - GUTCM}
\label{sec:3.1}

The GUTCM dataset includes data from 58 individuals diagnosed with SCD (39 females and 19 males, average age 65.24 ± 5.56 years), 89 individuals with MCI (63 females and 26 males, average age 65.31 ± 6.70 years), and 67 age-matched health volunteers (43 females and 24 males, average age 64.48 ± 5.73 years) as the healthy control (HC) group. All participants provided informed consent, and the study was approved by the institutional ethics committee.
The fMRI data were acquired using a 3.0 Tesla Siemens scanner, employing axial scans with a layer thickness of 5 mm, a Time Repetition (TR) of 2000 ms, an Echo Time (TE) of 30 ms, a flip angle of 90°, a Field of View (FOV) of 240 mm × 240 mm, a resolution of 64 × 64, and 31 layers. The DTI data were obtained using a standard echo-planar imaging functional head coil, with a layer thickness of 3 mm, TR of 6800 ms, TE of 93 ms, a flip angle of 90°, FOV of 240 mm × 240 mm, a resolution of 256 × 256, and 46 layers.
Incorporating two different early stages (SCD and MCI) in our GUTCM dataset aids in validating the \textbf{\textit{HA-HI}} framework's early diagnostic efficacy.

\vspace{-3mm}
\subsection{Dataset from the public resource - ADNI}
\label{sec:3.2}

In this study, we included 96 individuals with MCI (41 females and 55 males, average age 73.4 ± 7.07 years) from the ADNI dataset. Additionally, 66 age-matched normal control (NC) were selected from the ADNI repository as the NC group (38 females and 28 males, average age 76.7 ± 6.47 years). 
The imaging data were acquired using a variety of scanners from Siemens and GE, characterized by a 2000 ms TR and a 30 ms TE, and comprising either 31 or 48 layers. 

\vspace{-3mm}
\subsection{Data preprocessing}
\label{sec:3.3}

The raw fMRI data underwent a five-step preprocessing routine using the SPM12 toolbox \cite{tzourio2002automated}. These steps included: 1) slice-timing correction, 2) head motion estimation and correction, 3) intra-subject registration, 4) co-registration, and 5) regression-based outlier removal. In the domain of connectivity analysis, the Dosenbach164 atlas was employed to extract Regions of Interest (ROIs) for reconstructing functional connectivity, encompassing both SFC and DFC patterns. Given the temporal resolution of the MRI scanner, we obtained a total of 160 dynamic temporal points in DFCs, with each point representing a functional connectivity frame. Additionally, the ALFF was extracted using the DPARSF tool \cite{yan2010dparsf} to illustrate regional characteristics.
For DTI data preprocessing, we used the PANDA toolbox \cite{cui2013panda} to extract fiber tracts, where the Anatomical Automatic Labeling template was employed to guide ROI segmentation. The fiber strength between two ROIs was estimated based on the number of fibers. This was then normalized by the average surface area between the white and gray matter of those brain regions to derive the input features.

\vspace{2mm}
\section{Experimental Setup}
\label{sec:4}


In our experiments, the GUTCM and ADNI datasets were divided into training, validation, and testing sets with an 8:1:1 ratio, using cross-validation for result assessment. Within our \textbf{\textit{HA-HI}} framework's \textit{DMHA} component, convolution kernels were set to a size of 3 and stride of 1, except for the first layer in each modality branch and the pyramid adaptive convolution pipeline, where the stride was 2 to aid in dimensional reduction. The number of filters in this pipeline increased progressively with the temporal scale, starting from 8 and doubling to 16, 32, and 64. The \textit{DDHI} component employed a 16-head single-layer attention mechanism. We trained the model with batches of 4 and a learning rate of $10^{-4}$, using the Adam optimization strategy. Experiments were carried out on a server with an NVIDIA 3090Ti GPU, utilizing the PyTorch.

\begin{table*}[h]
    \centering
    \caption{Comparison study between the proposed \textbf{\textit{HA-HI}} and baseline models based on the GUTCM dataset}
    \label{tab: Table1}
    \resizebox{\linewidth}{!}{
    \begin{tabularx}{\textwidth}{l*{9}{>{\centering\arraybackslash}X}}
        \toprule
         & $R_{10}$FCL & $R_{10}$AFCL & $R_{10}$T & $R_{10}$AT  & $R_{18}$FCL & $R_{18}$AFCL & $R_{18}$T & $R_{18}$AT & \textbf{\textit{HA-HI}} \\
        \midrule
        \textbf{SCD/HC} \\
        Accuracy & 0.779 & 0.791 & 0.851 & 0.895 & 0.766 & 0.912 & 0.896 & 0.779 & \textbf{0.950} \\
        Recall & 1.000 & \textbf{1.000} & 1.000 & 0.862 & 1.000 & 0.887 & 0.946 & 1.000 & 0.950 \\
        Precision & 0.779 & 0.791 & 0.841 & \textbf{1.000} & 0.766 & 1.000 & 0.936 & 0.779 & 0.950 \\
        F1-score & 0.873 & 0.876 & 0.909 & 0.924 & 0.865 & 0.937 & 0.934 & 0.873 & \textbf{0.950} \\
        \midrule
        \textbf{MCI/SCD} \\
        Accuracy & 0.687 & 0.750 & 0.688 & 0.562 & 0.625 & 0.688 & 0.625 & 0.656 & \textbf{0.889} \\
        Recall & 0.700 & 0.694 & 0.562 & \textbf{1.000} & 0.338 & 0.450 & 0.613 & 0.650 & 0.800 \\
        Precision & 0.699 & 0.791 & 0.852 & 0.562 & 0.750 & 1.000 & 0.750 & 0.667 & \textbf{1.000} \\
        F1-score & 0.689 & 0.73 & 0.667 & 0.718 & 0.440 & 0.604 & 0.650 & 0.653 & \textbf{0.889} \\
        \midrule
        \textbf{MCI/HC} \\
        Accuracy & 0.719 & 0.562 & 0.545 & 0.656 & 0.562 & 0.697 & 0.545 & 0.562 & \textbf{0.812} \\
        Recall & 0.889 & 1.000 & 1.000 & 0.562 & 1.000 & 1.000 & 1.000 & \textbf{1.000} & 0.900 \\
        Precision & 0.695 & 0.562 & 0.545 & 0.729 & 0.562 & 0.664 & 0.562 & 0.606 & \textbf{0.818} \\
        F1-score & 0.775 & 0.718 & 0.694 & 0.624 & 0.718 & 0.732 & 0.718 & 0.718 & \textbf{0.857} \\
        \midrule
        \textbf{MCI/SCD/HC} \\
        Accuracy & 0.638 & 0.574 & 0.652 & 0.516 & 0.532 & 0.489 & 0.625 & 0.457 & \textbf{0.700} \\
        Recall & 0.962 & 1.000 & 0.707 & 1.000 & \textbf{1.000} & 1.000 & 0.710 & 0.800 & 0.950 \\
        Precision & 0.749 & 0.656 & \textbf{0.942} & 0.617 & 0.632 & 0.612 & 0.900 & 0.606 & 0.850 \\
        F1-score & 0.831 & 0.779 & 0.791 & 0.741 & 0.754 & 0.745 & 0.794 & 0.663 & \textbf{0.880} \\
        \bottomrule
    \end{tabularx}}
\end{table*}

\section{Experimental Results}
\label{sec:5}



\subsection{Comparison Study}
\label{sec:5.1}

In the subsequent sections, we conduct a comparative analysis of our proposed \textbf{\textit{HA-HI}} framework against other existing methodologies, utilizing both our GUTCM dataset (\textit{refer to Section \ref{sec:5.1.1}})) and the publicly available ADNI dataset (\textit{refer to Section \ref{sec:5.1.2}}).

For this comparison, we selected state-of-the-art (SOTA) methods \cite{chen2023orthogonal,jeon2020enriched,dong2021integration,lei2021auto}  that have been previously applied to the ADNI dataset for identifying MCI  from HC patterns to benchmark against \textbf{\textit{HA-HI}}.
Additionally, to validate the distinct components of our proposed method, we constructed baseline models using existing algorithms that share architectural similarities with the \textbf{\textit{HA-HI}} framework. 
Specifically, ResNet10 \cite{szegedy2017inception} and ResNet18 \cite{szegedy2017inception} were chosen for their structural resemblance to the basic blocks in the \textit{DMHA} component of \textbf{\textit{HA-HI}}. 
Moreover, attention-based modules \cite{woo2018cbam} and transformer blocks with varying numbers of self-attention heads were included due to their mechanisms comparable to the dual-modality interactions in the \textit{DDHI} component of \textbf{\textit{HA-HI}}. In detail, the ResNet architecture was utilized as the backbone for initial feature extraction, followed by the modules mentioned above for final classification. Therefore, the baseline models were constructed in four configurations: \textbf{(1)} Directly connecting a ResNet model to a Fully Connected Layer (FCL) ($R_{*}$FCL); \textbf{(2)} Applying an attention mechanism to channels before forwarding to the FCL for classification ($R_{*}$AFCL); \textbf{(3)} Directly connecting a ResNet model to the transformer blocks ($R_{*}$T); \textbf{(4)} Using features with applied channel attention as input for the transformer blocks ($R_{*}$AT).

\textit{
For clarity, in the subsequent subsections, the abbreviation '$R_{*}$' will be specified as '$R_{10}$' for ResNet10 and '$R_{18}$' for ResNet18. The abbreviation 'T' is specifically defined as 'T-S' for configurations with 8 heads and 'T-L' for 32 heads, illustrating the impact of having fewer or more attention heads compared to the standard configuration of 16 heads. 
}

\subsubsection{Baseline comparison on the GUTCM dataset}
\label{sec:5.1.1}

We quantitatively evaluated the \textbf{\textit{HA-HI}} framework through binary (SCD/HC, MCI/SCD, MCI/HC) and ternary (MCI/SCD/HC) classification tasks, as detailed in Table \ref{tab: Table1}. For the ternary task, individuals with cognitive impairment (MCI and SCD) were considered as positive samples for the calculation of the recall metric. While our methods did not uniformly outperform the baseline models across all metrics, \textbf{\textit{HA-HI}} consistently exhibited superior performance in terms of accuracy and F1 score across all four tasks. This consistent outperformance demonstrates \textbf{\textit{HA-HI}}'s robust capability in balancing the recognition of different cognitive patterns, which contributes to its robustness across various classification tasks and bolsters its generalizability when being applied to other datasets. Interestingly, the most notable results were observed in the SCD/HC task, indicating the model better recognizes the synergy effect when existing weaker cognitive decline.

\begin{table*}[h]
    \centering
    \vspace{-3mm}
    \caption{Comparative study of \textbf{\textit{HA-HI}} against baseline models and SOTA methods based on the public ADNI dataset}
    \label{tab:Table2}
    \resizebox{\linewidth}{!}{
    \begin{tabular}{ >{
     \centering\arraybackslash}p{3cm}>{\raggedright\arraybackslash}p{2.5cm}>{\centering\arraybackslash}p{02.5cm}>{\centering\arraybackslash}p{2.5cm}>{\centering\arraybackslash}p{2.5cm}>{\centering\arraybackslash}p{2.5cm}
    }
    \toprule
    \multicolumn{2}{c}{Methods} & Accuracy & Recall & Precision & F1-score\\
    \midrule
    \multirow{5}{*}{$R_{10}$} & FCL & 0.690 & 0.830 & 0.680 & 0.720\\
     & AFCL & 0.590 & 0.900& 0.640 & 0.710\\
     & AT &0.730 & 0.750 & 0.780 & 0.765\\
     & AT-S &0.586 & 1.000 & 0.586 & 0.738\\
     & AT-L &0.656 & 0.650 & 0.667& 0.658\\
     & T & 0.730 & 0.750 & 0.780 & 0.720 \\
     & T-S & 0.500 & 0.800 & 0.500 & 0.600 \\
     & T-L & 0.656 & 0.650 & 0.667& 0.730 \\
    \cmidrule{2-6}
    \multirow{6}{*}{$R_{18}$} & FCL & 0.600 & 0.800 & 0.600 & 0.670 \\
     & AFCL & 0.600 & 0.400 & 0.400 & 0.400 \\
     & AT & 0.600 & 1.000 & 0.500& 0.730 \\
     & AT-S & 0.600 & 1.000 & 0.500& 0.730 \\
     & AT-L & 0.600 & 1.000 & 0.600& 0.750\\
     & T & 0.600 & 0.600& 0.600& 0.600 \\
     & T-S & 0.400 & 0.400 & 0.400 & 0.400 \\
     & T-L & 0.600 & \textbf{1.000} & 0.600& 0.653 \\
    \midrule
    OLFG \cite{chen2023orthogonal} & 0.671 & 0.697 & /& / \\
    ST-ED \cite{jeon2020enriched} & 0.705 & 0.729 & /& /\\
    FC-DNN \cite{dong2021integration} & 0.784 & / & /& / \\
    AWCMT \cite{lei2021auto} & 0.822 & 0.829 & /& 0.806 \\
    \textbf{\textit{HA-HI}} & \textbf{0.825} & 0.900 & \textbf{0.780} & \textbf{0.836}\\
    \bottomrule
    \end{tabular}
    }
\end{table*}

\subsubsection{State-of-the-Art comparison on the ADNI Dataset}
\label{sec:5.1.2}

Our proposed \textbf{\textit{HA-HI}} exhibits consistent stability when validated on the ADNI dataset, which comprises participants of different ethnic backgrounds compared to the GUTCM dataset and includes data collected from a variety of devices. The summarized results are presented in Table \ref{tab: Table2}. 


We observed that complexity feature encoders, such as {$R_{10}$}FCL/AFCL and {$R_{18}$}AT/T-L, when simply stacked in blocks, exhibit a significant bias towards MCI patterns, compromising generalizability in HC subjects, as indicated by recall and precision metrics. This underscores the need for systematic and rational integration in model architecture, a topic we delve into in Section \ref{sec:5.2} through a thorough examination of each component in the \textbf{\textit{HA-HI}} framework.


Additionally, we investigated the effect of varying the number of attention heads within transformer blocks to refine the models' ability to distinguish between MCI and HC. Our findings reveal that a model's performance does not linearly correlate with the number of self-attention heads. Identifying an optimal number of heads, which we found to be 16 for datasets similar in scale to ours, is crucial for balancing model capacity and data scale.

To benchmark against existing methods that have been applied to the ADNI dataset for the same task, we compared our results with four distinct works, including:
(1) an orthogonal latent space learning with feature weighting and graph learning model \cite{chen2023orthogonal}, termed as OLFG; 
(2) a spatiotemporal feature-based encoder-decoder framework \cite{jeon2020enriched}, termed as ST-ED;
(3) a functional connectivity-based deep neural network \cite{dong2021integration}, termed as FC-DNN;
(4) an auto-weighted centralized multi-task learning framework \cite{lei2021auto}, termed as AWCMT;

The \textbf{\textit{HA-HI}} notably outperformed other models, owing to its ability to co-analyze dual-domain features. This enhanced performance benefits from its hierarchical structure in the \textit{DMHA} component, enabling effective dual-modality alignment, and its  hierarchical structure in the \textit{DDHI} component, adeptly fusing high-level features from fine-grained to global levels.

\vspace{-3mm}
\subsection{Ablation study}
\label{sec:5.2}


\subsubsection{Ablation Study on Feature Modalities}
\label{sec:5.2.1}

The enhanced overall performance with dual-modal MRI, detailed in Table \ref{tab: Table3}, aligns with our hypotheses. Notably, the highest recall was observed in DFC analysis, with a decline in recall from dynamic to static features and from connectivity to regional domains. Conversely, precision increased under these conditions. Structural regional features, while yielding lower recall, maintained accuracy comparable to functional features, suggesting higher specificity.
These findings indicate: 1)Functional network disruptions are significant in early cognitive impairment, serving as vital indicators, in line with existing literature. 2) Functional features (DFC, SFC, ALFF) are integral to decision-making: DFC enhances sensitivity to abnormalities (as seen in high recall), while ALFF improves accuracy in identifying positive cases (evident in high precision). 3) Regional features provide valuable complementary insights to connectivity features, aiding in reducing false positives.

\begin{table}[t]
\caption{Ablation study on MRI dual-modalities: Quantitative analysis across different branches in the \textbf{\textit{HA-HI}} framework using the DFC, SFC, ALFF, and FA modalities, compared with the combined use of all modalities}
\label{tab: Table3}
\resizebox{\linewidth}{!}{
    \begin{tabularx}{\columnwidth}{XXXXX}
    \toprule
    & Accuracy & Recall & Precision & F1-score \\
    \midrule
    DFC & 0.580 & \textbf{1.000} & 0.610 & 0.740\\
    SFC & 0.690 & 0.930 & 0.680 & 0.760 \\
    ALFF & 0.660 & 0.850 & 0.700 & 0.730 \\
    FA & 0.650 & 0.740 & 0.660 & 0.600 \\
    \textbf{Dual-MRI} & \textbf{0.825} & 0.900 & \textbf{0.780} & \textbf{0.836} \\
    \bottomrule
    \end{tabularx}}
\end{table}

\subsubsection{Ablation Study on Model Architectures}
\label{sec:5.2.2}

To assess the necessity of each \textbf{\textit{HA-HI}} component, we evaluated the model on the ADNI dataset by systematically removing alignments (\textit{DMHA: TSA|DSA|FSA}) and interactions (\textit{DDHI: FI|GI}), as shown in Table \ref{tab: Table4}. We first removed the cross-domain attention modules (\textit{GI} in \textit{DDHI}), followed by bypassing \textit{FI} in \textit{DDHI}, directly feeding \textit{DMHA} outputs to the residual projector. We then eliminated the contrastive restrictions in \textit{FSA} and \textit{DSA}, and the pyramid adaptive convolution pipeline in \textit{TSA}, to evaluate the impact of hierarchical alignments.

\begin{figure*}[h]
\centering
    \includegraphics[scale=.41]{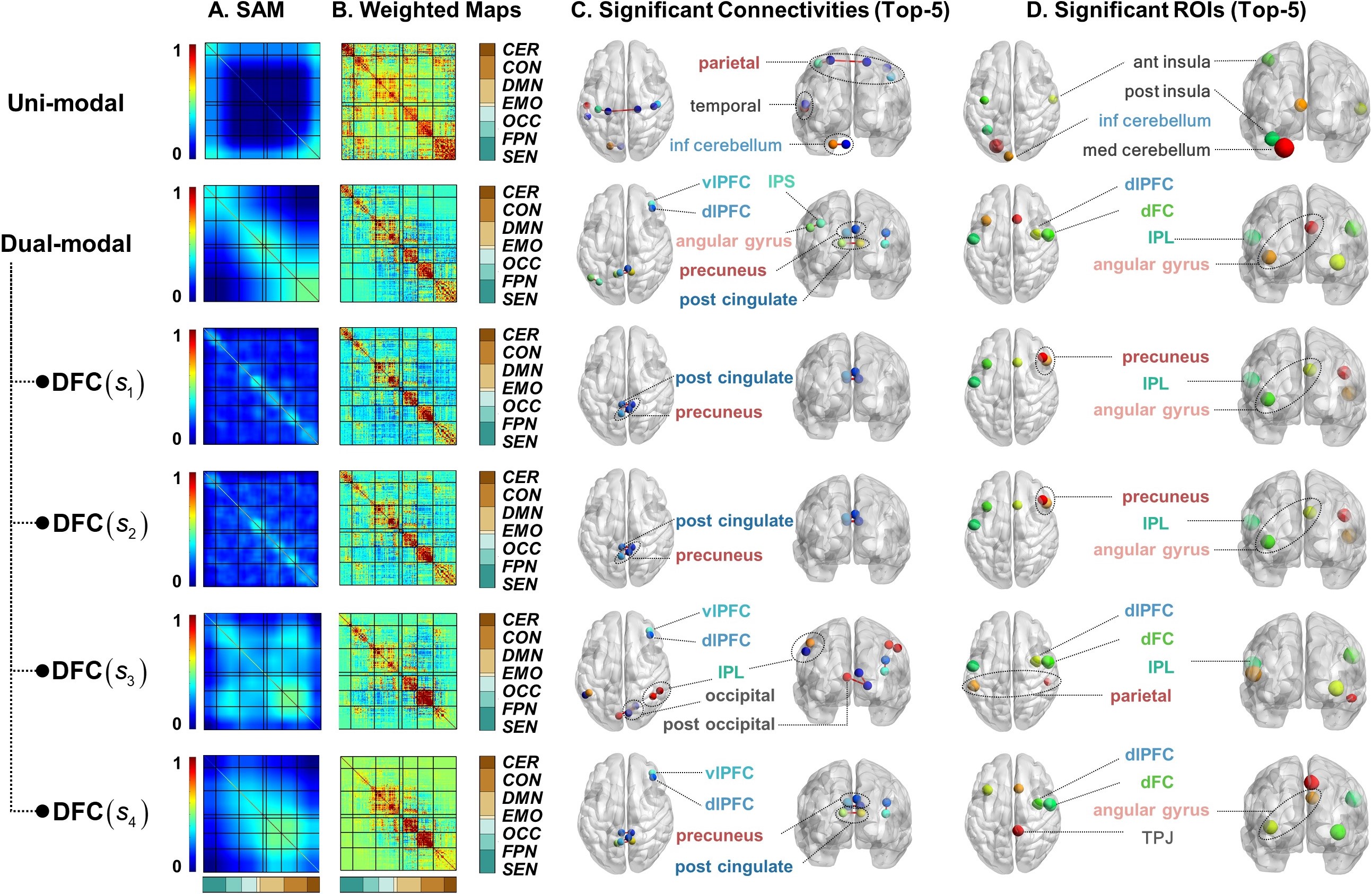}
    \caption{
    MRI synergy effects reflected by multi-scale DFC features. A) Activation maps. B) Weighted feature maps, calibrated with the activation maps. C) Top five significant connectivities. D) Five key regions with notable ROI-wise connectivity strength. Networks represented include CER (cerebellum network), CON (cingulo-opercular network), DMN (default mode network), EMO (emotional network), OCC (occipital network), FPN (frontoparietal network), and SEN (sensorimotor network)
    }
    \label{fig: Synergy Effects-DFC}
\end{figure*}

\begin{table*}[b]
    \caption{Ablation study on model architectures: Quantitative analysis of components in the \textbf{\textit{HA-HI}} framework under hierarchical removal scenarios}
    \label{tab: Table4}
    \begin{tabularx}{\textwidth}{*{9}{>{\centering\arraybackslash}X}}
     \toprule
     \multicolumn{3}{c}{\textit{DMHA}} & \multicolumn{2}{c}{\textit{DDHI}} & \multirow{2}{*}{Accuracy} & \multirow{2}{*}{Recall} & \multirow{2}{*}{Precision} & \multirow{2}{*}{F1-score}\\
     \cmidrule(l){1-3} \cmidrule(l){4-5}
     TSA & DSA & FSA & FI & GI \\
     \midrule
     \checkmark &\checkmark &\checkmark &\checkmark &\checkmark & \textbf{0.825} & 0.900 & 0.780 & \textbf{0.836}\\
     \midrule
     \checkmark &\checkmark &\checkmark &\checkmark &\ding{55} & 0.750 & \textbf{1.000} & 0.750 & 0.857\\
     \midrule
     \checkmark &\checkmark &\checkmark &\ding{55} &\ding{55} & 0.724 & 0.724 & 0.690 & 0.707\\ 
     \midrule
     \checkmark &\checkmark &\ding{55} &\ding{55} &\ding{55} & 0.690 & 0.655 & 0.621 & 0.638\\
     \midrule
     \checkmark &\ding{55} &\ding{55} &\ding{55} &\ding{55} & 0.607 & 0.571 & 0.548 & 0.559\\
     \midrule
     \ding{55} &\ding{55} &\ding{55} &\ding{55} &\ding{55} & 0.517 & 0.238 & \textbf{0.816} & 0.369\\
     \bottomrule
    \end{tabularx}
\end{table*}

The gradual decrease in accuracy and F1 score, as detailed in Table \ref{tab: Table4}, underscores each component's importance in \textbf{\textit{HA-HI}}. Notably, omitting the final connectivity-regional interaction maximized recall, while the lowest recall and highest precision occurred when using DFC, SFC, ALFF, and FA as inputs without any alignments or interactions. This implies that while dual-modal MRI features can challenge model generalizability, leading to missed detections, hierarchical alignments significantly counter these challenges. Additionally, attention-based cross-modality interactions, particularly in a hierarchical structure, enhance disease detection efficiency and aid in balancing the identification of both positive and negative samples.

\subsection{The Synergy Effects in Dual-modal MRI}
\label{sec:5.3}
\vspace{-3mm}



\subsubsection{Connectivity domain: uni-modal vs. dual-modal scenarios}
\label{sec:5.3.1}

The activation maps in Fig. \ref{fig: Synergy Effects-DFC}A, obtained using the \textit{SAM} technique, demonstrate a bias in the uni-modality model towards peripheral regions, unlike the dual-modal framework. This bias could lead to decreased precision in positive detections (see 'DFC' row in Table \ref{tab: Table3}). In contrast, the dual-modal approach shifts focus to broader network connectivities. Furthermore, multiscale analysis effectively merges activation maps from various temporal scales for a more comprehensive and balanced focus.
Using weighted DFC feature maps (Fig. \ref{fig: Synergy Effects-DFC}B), calibrated by activation maps (Fig. \ref{fig: Synergy Effects-DFC}A), we visualized and compared the top five significant connectivities (Fig. \ref{fig: Synergy Effects-DFC}C) under various observation scenarios. 
Additionally, we evaluated the importance of each ROI in the connectivity domain by averaging associated connectivities, showcasing the top five ROIs in Fig. \ref{fig: Synergy Effects-DFC}D. A comparison between uni-modal and dual-modal scenarios indicates a shift in focus from cerebellar to DMN and FPN networks in dual-modal MRI, aligning with their known cognitive function relevance \cite{chen2023default}. This adjustment is achieved through alignments across multiple temporal scales and dynamic-static patterns in the \textit{TSA} and \textit{DSA} modules of the \textbf{\textit{HA-HI}} framework. Significantly, in the dual-modal scenario, the notable regions are influenced by various temporal scales. Specifically, fine-grained temporal scale characteristics (e.g., $DFC(s_0)$ and $DFC(s_1)$) predominantly manifest in the DMN, while coarser-grained DFC features (e.g., $DFC(s_2)$ and $DFC(s_3)$) more strongly indicate abnormalities in the FPN. These observations underscore the synergy effects among multi-scale DFCs, which are analyzed and integrated through the pyramid adaptive convolution pipeline in the \textit{TSA} module, potentially enhancing the accuracy of detection.

\begin{figure}[ht]
\centering
    \includegraphics[scale=.45]{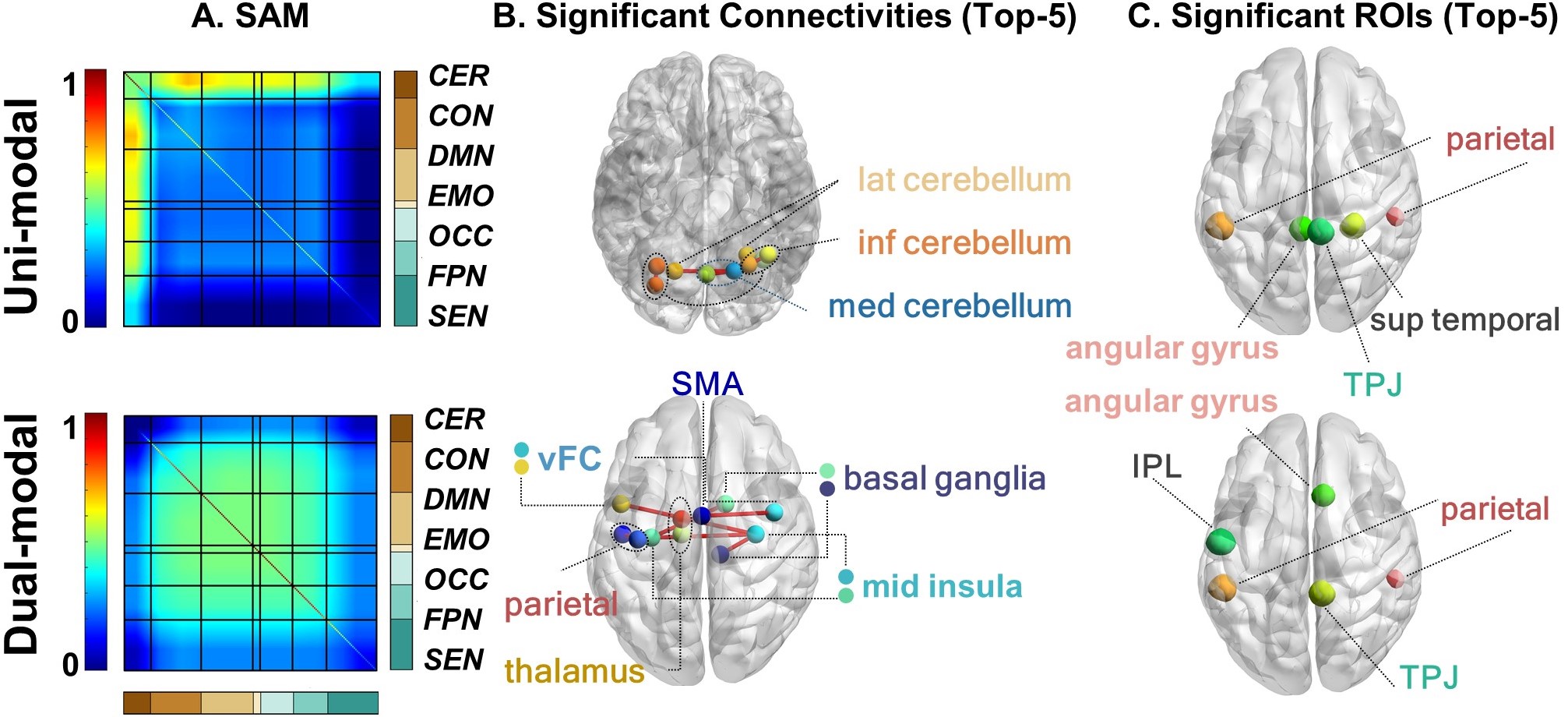}
    \caption{MRI synergy effects from an SFC Perspective. A) Activation maps from the \textit{SAM} technique. B) Top five connectivities impacted by cognitive impairment. C) Five regions with notable abnormal ROI-wise connectivity strength.
    \vspace{-3mm}
    }
    \label{fig: Synergy Effects-SFC}
\end{figure}

The SFC results closely mirror those from the DFC perspective, with the uni-modal framework showing a bias towards cerebellum network connectivities. However, in the dual-modal scenario (Fig. \ref{fig: Synergy Effects-SFC}A), there is a noticeable shift in focus towards the DMN, FPN, and CON networks. This shift underlines the effectiveness of our \textbf{\textit{HA-HI}} framework in combining multi-scale DFCs and SFC to highlight abnormal connectivities in cognitive-related networks such as DMN, FPN, and CON \cite{chen2023default}. Notably, a comparison between the DFC and SFC perspectives in the dual-modal setup reveals that DMN and FPN connectivities are predominantly influenced by dynamic features, whereas CON abnormalities are more distinct in static features.

\subsubsection{Regional domain: uni-modal vs. dual-modal scenarios}
\label{sec:5.3.2}

\begin{figure}[h]
\centering
\includegraphics[scale=.45]{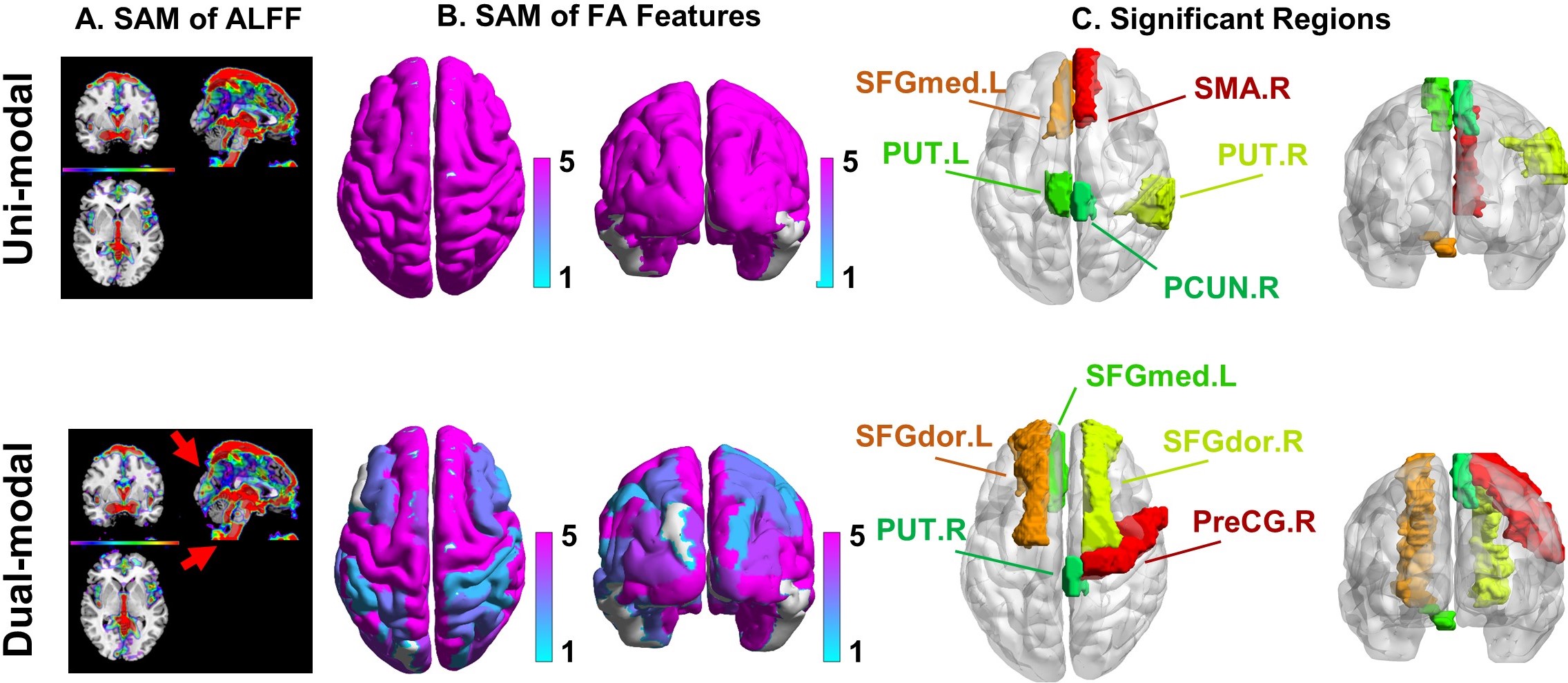} 
    \caption{  
    Regional domain synergy effects in dual-modal MRI. A) Significant areas as reflected by ALFF features in fMRI. B) Activation maps indicated by FA features in DTI. C) The top five brain regions exhibiting significant regional deterioration.
    }
\label{fig: Synergy Effects-Region}
\end{figure}

In fMRI, regional features measured by ALFF indicate spontaneous brain activity. As shown in Fig. \ref{fig: Synergy Effects-Region}A, the parietal lobe, occipital lobe, cerebellum, thalamus, and hippocampal gyrus are most affected by cognitive impairment, with the hippocampal gyrus being a key cognitive region \cite{yang2018gradual}. The activation patterns in both uni-modal and dual-modal scenarios are largely consistent, with a notable difference in the dual-modal framework's broader focus on the cerebellum and parietal lobe. This variation may be influenced by other modalities, such as the FA feature in DTI affecting the parietal lobe and connectivity features in fMRI impacting the cerebellum. These observations suggest that the \textbf{\textit{HA-HI}} framework could enhance the generalizability of learned representations.


The significant regions identified based on the FA features indicate evidence of the most degenerated fibers associated with these regions. The visualization of the top five regions (Fig. \ref{fig: Synergy Effects-Region}B-C) with significant alterations in cases of cognitive decline shows a marked trend of shifting focus from the dorsal side to the ventral side of the brain.
The dual-modal framework intensifies the focus on the frontal and prefrontal lobes. {These structural changes in these brain regions are believed to be closely associated with cognitive decline \cite{schmithorst2005cognitive}}.



\vspace{-6mm}
\section*{Conclusion}

In this study, we hypothesize that aligning and fusing structural-functional regional features, static and dynamic functional connectivity, as well as dynamic functional networks across multiple temporal scales, could enhance the detection of cognitive impairment in MRI analyses. 
To test this hypothesis, we introduce the novel \textit{\textbf{HA-HI}} method. This approach synergizes fMRI and DTI data by hierarchical alignments and interactions. Specifically, feature alignment is achieved via a \textit{DMHA} module, operating on a horizontal hierarchy, while cross-domain interactions are facilitated by a \textit{DDHI} module, employing a vertical hierarchy that extends from fine-grained to global levels.
\textit{\textbf{HA-HI}} was evaluated on GUTCM (a local hospital dataset) and ADNI (a public resource) datasets, showing superior quantitative performance over baseline and SOTA methods.
Qualitatively, we contrasted the core features identified by the uni-modal strategy with those highlighted by our approach. 
For this, we developed a \textit{SAM} technique, which reveals significant functional networks and brain regions impacted by cognitive impairment. Our findings indicate that cognitive impairment most severely affects functional networks such as DMN, FPN, and CON. 
Additionally, our method emphasizes the importance of the frontal and prefrontal lobes at the functional level, and the thalamus and hippocampal gyrus at the structural level. In conclusion, our deep-learning diagnosis model, accompanied by an interpretable tool, offers valuable insights into multi-modal MRI analysis technically, contributing to the theoretical development in the study of cognitive decline.

\section{Acknowledgement}

\thanks{
This research was supported by the National Natural Science Foundation of China (Grants 62306089, 32361143787, 82102032), the China Postdoctoral Science Foundation (Grants 2023M730873, GZB20230960), and the Guangxi Natural Science Foundation (Grant No. 2023GXNS- FBA026073). (Corresponding authors: Zhenxi Song, Demao Deng and Zhiguo Zhang. Xiongri Shen and Zhenxi Song contributed equally to this work.)}

\thanks{Xiongri Shen, Zhenxi Song, Min Zhang, and Zhiguo Zhang are with the Department of Computer Science and Technology, Harbin Institute of Technology, Shenzhen, 518055, China (email: xiongrishen@stu.hit.edu.cn, songzhenxi@hit.edu.cn, zhangmin2021@hit.edu.cn, zhiguozhang@hit.edu.cn).}

\thanks{
Linling Li is with the Guangdong Key Laboratory of Biomedical Measurements and Ultrasound Imaging, School of Biomedical Engineering, Shenzhen University Medical School, Shenzhen University, Shenzhen 518060, China (lilinling@szu.edu.cn).}

\thanks{
Honghai Liu is with the State Key Laboratory of Robotics and Systems, Harbin Institute of Technology, Shenzhen 150001, China, and also with Peng Cheng Laboratory, Shenzhen 518000, China (e-mail: honghai.liu@hit.edu.cn).
}

\thanks{
Demao Deng, Yichen Wei, Lingyan Liang is with the Department of Radiology, The People’s Hospital of Guangxi Zhuang Autonomous Region, Guangxi Academy of Medical Sciences. Nanning, China (email: demaodeng@163.com, 316644690@qq.com, lianglingyan163@126.com).}



\clearpage

\end{document}